\documentstyle[preprint,aps]{revtex}

\begin{document}
\title{Helical Ribbons as Isometric Textures}
\author{M.-F. Achard$^{1}$, \ M. Kleman$^{2}$\thanks{%
author for correspondence; e-mail: maurice.kleman@mines.org}, \ Yu. A.
Nastishin$^{1,3}$, \ H.-T. Nguyen$^{1}$}
\address{$^{1}$Centre de Recherche Paul Pascal, CNRS, Universit\'{e}\\
Bordeaux 1, Avenue Schweitzer, 33600 Pessac, France\\
$^{2}$Laboratoire de Min\'{e}ralogie-Cristallographie de Paris, Universit%
\'{e} Pierre-et-Marie-Curie,\\
Case 115, 4 place Jussieu, 75252 Paris c\'{e}dex 05, France\\
$^{3}$Institute for Physical Optics, 23 Dragomanov str., Lviv, 79005 Ukraine}
\date{\today }
\maketitle

\begin{abstract}
Deformations that conserve the parallelism and the distances --between
layers, in smectic phases; between columns, in columnar phases-- are
commonplace in liquid crystals. The resulting deformed textures have the
same mass density as in the ground state (an expected property in a liquid)
and are at the same time isogonic and isometric, which imposes specific
geometric features. The corresponding order parameter singularities extend
over rather large, macroscopic, distances, {\it e.g}., cofocal conics in
smectics. This well-known picture is modified when, superimposed to the 1D
or 2D periodicities, the structure is helical. Isogony is no longer the
rule, but isometry (and mass density) can be preserved. This paper discusses
the case of a medium whose structure is made of 1D modulated layers (a
lamello-columnar phase), assuming that the modulations rotate helically from
one layer to the next. The price to pay is that any isometric texture is
necessarily frustrated; it consists of layers folded into a set of parallel
helicoids, in the manner of a screw dislocation (of macroscopic Burgers
vector), the modulations being along the helices, i.e. double-twisted. The
singularity set is made of two helical disclination lines. We complete this
geometric analysis by a crude calculation of the energy of a helical ribbon.
It is suggested that the helical ribbons observed in the B7 phase of
banana-like molecules are such isometric textures. As a side result, let us
mention that the description of double-twist, traditionally made in terms of
a partition of the director field into nested cylinders, could more than
often be profitably tested against a partition into nested helicoids.
\end{abstract}


{\footnotesize So full of shapes is fancy,}

{\footnotesize That it alone is high fantastical.}

{\footnotesize {\it Twelfth Night }[I, 1, lines 14-15]}

\section{\protect\bigskip Introduction}

Consider a bent solid crystal (Fig. 1a); it is well known since \cite{Nye}
that the bending stresses can be relaxed by a set of dislocations of density 
$1/bR$, where R is the radius of curvature and b the Burgers vector. These
dislocations (in this example, edge dislocations) are spaced at a distance $%
\approx \sqrt{bR}$. Under annealing they assemble in the shape of walls;
this process is known by the name of {\it polygonization}\cite%
{JFriedel,Nabarro}. In this geometry (Fig. 1b) the layers of atoms parallel
to the Burgers vector keep practically unstrained, and the largest
contribution to the strain energy comes from the cores of the dislocations.
The repeat distance in a direction perpendicular to the Burgers vector is
conserved and the atoms have almost the same density along the curved layer
as in the unbent crystal; the curvature can be large. The same geometry
transposed to a layered liquid, a smectic A (SmA), shows that the curvature
deformation is compensated by liquid relaxation along the layers, Fig. 1c %
\cite{MKlemanLavrentovich}. Transposed to a columnar phase, the columns
curvature is such that their equidistance and parallelism are conserved,
which implies that they are orthogonal to a family of planes; the 2D network
of their intersections with any of those planes is unchanged compared to the
undeformed state \cite{MKlemanLavrentovich}. In a sense it is still feasible
to speak of dislocations in lamellar and columnar phases, but they now have
a vanishing Burgers vector, their density is infinite, and there is no
analogue to polygonization. The energy of deformation is pure curvature
energy; the mass density is conserved. Such deformations, we shall call them 
{\it isometric}, in the sense that they do not affect the 'solid' component
of the ordered liquid, i.e. the distance between lamellae or between
columns. In the columnar case, they also are {\it isogonic}.

\[
\text{{\LARGE Fig.1}} 
\]

Isometric deformations are commonplace in 1D and 2D liquid crystals. It is
an easy matter to show that they must dominate on scales larger than $\sqrt{%
\frac{K}{B}}$, where $K$ is a typical Frank curvature modulus and $B$ a
typical inverse compressibility, {\it i.e.} on scales larger than a few
repeat distances \cite{MKlemanLavrentovich}. They form characteristic
textures that carry singularity sets. The case of smectics has been known
for more than a century \cite{GFriedel}; the relevant isometric textures are 
{\it focal conic domains}, which are singular on a pair of conjugate conics,
the layers take the shape of {\it Dupin Cyclides}. In columnar phases the
isometric textures are {\it developable domains} \cite{Kleman1,Bouligand},
which are singular on a developable surface; the columns can be partitioned
(in an infinity of ways) into families of parallel lines belonging to {\it %
Monge Surfaces}. We refer the reader to Ref. \cite{MKlemanLavrentovich},
chapter 10, for a full description. It is a remarkable fact that these
objects, whose geometry is so precisely defined, extend over mesoscopic
scales, quite often of the order of a mm, {\it i.e.} involving 10$^{5}$ to 10%
$^{6}$ strictly parallel layers in a SmA. This indicates the extension of
the isometric principle in ordered liquids.

Let us come now to a newcomer in the realm of mesomorphic structures, namely
the B7 phase. It is widely accepted that the structure of this phase
involves at the same time a smectic layering (Sm) and a one-dimensional
modulation (D) of the layers \cite{Walba} (we call `columns' the elements of
this modulation, in a metaphoric sense; we denote by $\overrightarrow{n}$
their unit tangent vector; the symbol D stands for `order of the columnar
type', as with discotic thermotropic molecules). It is believed that the B7
order is built upon a low-symmetry, chiral, molecular configuration \cite%
{Brand}, although the molecules are non-chiral. These molecules are planar
and bent-like, with a rather strong dipole moment along an axis of symmetry,
Fig. 2. Therefore domains of opposite chiralities coexist in the ground
state. But at a mesoscopic level, notwithstanding this {\it chiral}
character of the structure, the ground state does not exhibit {\it helicity}
(only Sm and D orders are simultaneously present). Observe that the lamellar
structure allows that the modulations of neighboring layers keep in
register, yielding thus close-packing and high mass density.

Were it limited to these two types of order, Sm+D, one would expect that the
textures of lower energy, {\it i.e.}, the isometric distortions, although
somewhat more subtle than the isometric distortions of sets of parallel 2D
liquid layers or of parallel 1D liquid columns, would not require an
analysis of a degree of difficulty much deeper, in the sense that no new
geometrical concept would be necessary \cite{MKleman2}.

A most conspicuous feature of the B7 phase is the presence of {\it helical
ribbons} (HR) that appear in the isotropic phase when the samples are cooled
down rapidly from the high temperature range of the isotropic phase \cite%
{Walba,Bedel,YuNastishin,Jakli,Coleman}. These nuclei clearly display a
double-helix texture, Fig. 3. Left and right HRs are in approximately equal
numbers. A linear relation between the pitch $b$ and the width $l$ has been
documented by several authors \cite{Walba,Jakli}. The model for HRs proposed
by Coleman \& {\it al}. \cite{Coleman} consists in two cylinders, made of
tightly packed cylindrical layers, twisted one about the other. This model
is in full agreement with the Sm+D structure of the B7 phase, but is not
isometric, in the sense that the twisting of the two cylinders hinders
close-packing.

\[
\text{{\LARGE Fig.2}} 
\]

There are in fact two types of HRs \cite{YuNastishin}. The S-HRs (slim) are
characterized by a rather perfect double helical geometry, a pitch/width $%
b/l $ ratio $\approx \pi $ (see Fig. 3), and a high rigidity; they do not
bend easily, which is related to the remarkable constancy of the pitch and
of the $b/l$ ratio all along each ribbon. The F-HRs (fat) are much more
disperse in pitch, shape, and width. We shall in this note expatiate only on
S-HRs, (simply noted HRs in the sequel). We present in the next section a
purely geometric model, which interprets HRs as isometric textures of the B7
phase; and explains satisfactorily well their double helical geometry and
the value of the ratio $b/l\cong \pi $. In particular, the model is such
that the modulation of any B7 twisted layer does match the modulation of the
neighboring twisted, parallel, B7 layers, as required by the isometric
principle.

\[
\text{{\LARGE Fig.3}} 
\]

The helical pitch $b$ varies somewhat from one nucleus to another, but about
a value that is rather strongly peaked, Fig. 4. Also the S-HRs transform in
a period of a few minutes into F-HRs. These features have led the authors of %
\cite{YuNastishin} to assume that S-HRs are specific textures of metastable
modifications of the B7 phase, which display helicity (H order), either left
or right, superimposed on Sm and D orders. We note them B7$^{\text{*}}$. The
axis of helicity is perpendicular to the layers and affects the columns. We
thus assume that the B7$^{\text{*}}$ helical states differ from B7 by the
existence of a finite pitch $p$ of the columns, $p$ being a material
constant. We show, in the framework of a very simple model, that the free
energy is minimized when $\varpi q\cong -1$ (where $\varpi =b/2\pi $, $\
q=2\pi /p$), {\it i.e.} when $b\cong -p,$ which is precisely what is
expected from the purely geometric model, and that this energy is less than
the energy of the model advocated in \cite{Coleman}.

\[
\text{{\LARGE Fig.4}} 
\]

The next section is devoted to a detailed account of our HR geometric model;
section III deals with an approached calculation of the HR energy, assuming
the existence of a metastable B7$^{\text{*}}$ phase.

\section{The Geometrical Model.}

\subsection{Giant Screw Dislocations in SmAs, the Chromosome of
Dinoflagellate, Double-Twist, and all that.}

Our isometric model is related in many ways to the three topics mentioned in
the title of this subsection. We briefly expose this relationship with these
topics, before (in the next subsection) giving a more detailed account of it.

Let us first summarize the main features which characterize a HR according
to our model: a HR is the inner part of a giant Burgers vector screw
dislocation; the layers are deployed along parallel and equidistant
helicoids whose common axis is the dislocation line and whose pitch is equal
to the Burgers vector; the columns are along parallel helices of the same
axis and belong to the helicoids. The singularity set of this isometric
texture is made of a double helix located at the periphery of the whole
texture. Fig. 5 is a mathematical illustration of the model; see also Fig. 8
of \cite{YuNastishin}.

\[
\text{{\Large Fig.5}} 
\]

Screw dislocations with giant Burgers vectors are documented in SmAs since
long \cite{Williams}; they are characterized by the fact that the core is
split into two helical singularities located at a distance $\varpi =b/2\pi $
from the axis of the screw, $b$ being its Burgers vector; this geometry is
visible under the optical microscope, $b$ being so large (typically a few $%
10\mu m$ in CBOOA, a classic thermotropic smectic; the Burgers vector is not
related to any material constant, but rather to the circumstances of the
experiment, very probably the thickness of the sample). This geometry can be
explained as follows \cite{Frank,MKlemanLavrentovich}. The SmA layers form a
set of {\it parallel}, {\it equidistant} surfaces (isometric principle!),
which are nested helicoids, stacked upon a central {\it ruled helicoid}%
\footnote{%
It does not come as a surprise that the same object is sometimes called
'right helicoid' in Anglo-Saxon countries (see for example \cite{Hilbert}),
and 'h\'{e}lico\"{\i}de {\it gauche} \`{a} plan directeur' by the French %
\cite{Darboux}; of course, in these expressions, neither the attributive
'right', nor the \'{e}pith\`{e}te '{\it gauche}' refer to handedness! We use
a third denomination.} whose axis is along the screw line; it has a right or
left hand, according to the case, analytically distinguished by the sign of $%
b$. A ruled helicoid is generated by a straight-line meeting a fixed axis at
a right angle and moving with a uniform helical motion about this axis; it
has a vanishing mean curvature:%
\begin{equation}
H=\frac{1}{2}(\sigma _{{\rm 1}}+\sigma _{{\rm 2}})=0\text{ .}
\label{eq:ZeroMinCurv}
\end{equation}

The two singularities are the cuspidal edges of the focal surfaces of the
set of helicoids; they act physically as two helical {\it disclinations} of
strength $k=1/2$ (see \cite{MKlemanLavrentovich}), about which the layers
fold. Outside the cylindrical region bound by the disclinations, the layers
adopt a quasi-planar geometry that is not in analytical continuity with the
helicoids inside, but are in register with the layers inside.

We argue that the B7$^{\text{*}}$ layers in the HR stack as those of the SmA
phase in the screw dislocation. The geometric properties of the layer
stacking impose that the disclination lines are scribed on a cylinder of
diameter $l=b/\pi $. This is precisely what one observes empirically in the
B7$^{\text{*}}$ case, taking for $l$ the measured diameter of the HR. But
there are differences. The HR is limited to the core region, inside the
cylinder on which the disclinations are scribed, whereas the SmA screw
dislocation extends outside the core region. Also, it is precisely the
helicity of the columns (no columns in the SmA phase, of course) that
stabilizes the HR. The screw dislocation is a {\it metastable} defect of the
SmA phase. The HR, on the other hand, see next section, is the texture of 
{\it lowest energy} of the frustrated metastable B7$^{\text{*}}$ phase.

The HRs bear also some resemblance with a model \cite{MKleman3} devised in
order to explain the conformation of the molecule of DNA in the chromosome
of Dinoflagellate, observed by Livolant \& Bouligand \cite{Livolant}. The
same observations also attracted the attention of J. Friedel, who concluded
to a much similar model \cite{JFriedel2}. As shown in \cite{Livolant}, the
DNA is cholesteric in this chromosome, but at the same time the isometric
principle, which applies here because the persistence length of the molecule
is not negligible compared to the cholesteric pitch, requires that the local
packing be as dense as possible. Ref\cite{MKleman3} discusses precisely the
possibility of the coexistence of cholesteric and 2D local order, which {\it %
a priori} are incompatible in the large, but, as it was shown, can be fitted
locally if the molecules can be partitioned into a family of parallel,
helical, surfaces containing their long direction. Here sits the analogy
with HR. But whereas the columns are along helices in a HR, it is believed
that the DNA molecules (but to shape fluctuations) are along a set of lines
orthogonal to these helices, namely a set of geodesics\footnote{%
This is not the densest packing; but it is the straightest configuration,
which is in accord with the long persistence length of DNA.}.

In B7$^{\text{*}}$ the columns sit along helices with approximately the
opposite pitch to the layers pitch; thus, they display the well-known blue
phase (BP) {\it double-twist} geometry. In the case of BPs, this geometry is
not related to the presence of any layers: the columns are in fact
liquid-like in the three directions, and the order parameter is locally
cholesteric. Double-twist is splayless and its bend energy is relatively
small, as long as the lateral size of the bundle of double-twisted molecules
does not exceed $\approx b/2$ (the double-twist geometry is frustrated). It
has been stressed \cite{MKleman4,Kleman5} that in ideal double-twist
geometry, the columns are equidistant, having the same property of
equidistance met {\it e.g.}, for the strands of a string. It is such a
property we expect that the columns of the B7$^{\text{*}}$ phase do satisfy,
with the result that the elastic energy is not prohibitive. How this
requirement relates to helical layers is discussed now. But whereas most
people think of the double-twist geometry as geometry of nested cylinders,
into which the columns are partitioned, the screw dislocation model drives
us to partition the HR geometry into helical surfaces. We shall come again
to the nature of this discrepancy, which is not as trivial as it might look
at first sight.

\subsection{Why Helicoids}

Helicoids possess a property that they share with surfaces of revolution:
they are invariant under a {\it one-parameter family of} \underline{{\it %
rigid motions}} \cite{Hilbert}, namely a continuous set of
translation-rotations about the helical axis (for helicoids), a continuous
set of rotations about the axis of revolution (for surfaces of revolution).
Of course the circular cylinder (limit case of a helicoid for $\varpi =0$)
and the sphere possess respectively two- and three-parameter families of
rigid motions, but we leave aside these limit cases, that are trivial. Less
trivial is the fact that surfaces of the type under consideration, namely
helicoids (resp. surfaces of revolution), contain two orthogonal families of
curves of which one family is the family of helices (resp. circles of
latitude) whereas the other one consists of geodesics (resp. meridians).
This is exactly the geometrical property we are looking for: if one installs
the columns of the B7$^{\text{*}}$ phase along the curves orthogonal to the
geodesics, then the columns stay at equal distances, because of the property
of {\it frontality} of the geodesics.

The demonstration of the existence of two orthogonal families of curves \cite%
{Hilbert} requires the recognition that the rigid motion invariance is also
invariance under a {\it one-parameter family of }\underline{{\it bendings}}.
A surface admits bending invariance if the following is true. Place, snug
against the surface, a piece of flexible but not stretchable foil; then it
can be displaced along the surface, at least along one direction, so that it
keeps snug against it without tearing, although it can change its shape. If
the surface admits a one-parameter family of bendings, then there are
curves, which cover the whole surface, such that the displacements along
those curves exhibit the above property of invariance. Because bending does
not change geodesic distance, any two curves of the family must be at a
constant geodesic distance. Therefore a geodesic line orthogonal to one of
the curve of the first family crosses at right angles all the curves of the
first family. Because bending keeps Gaussian curvature invariant, Gaussian
curvature is constant all along a curve of the first family. Gaussian
curvatures measured along two curves of the first family are generically
different. Therefore the curves of the first family do not cross; they form
a family without singular points.

This first family of lines is a set of helices. It can indeed be shown \cite%
{Hilbert} that surfaces admitting a one-parameter family of bendings can be
brought into the form of general helicoids or of surfaces of revolution.
Most evidently the curves of the first family are the helices (resp. the
circles of latitude), and no other curves. Therefore the condition of
equidistance (more precisely: of equal geodesic distance) forces the columns
to sit along the helices of general helicoids (resp. along the circles of
latitude of surfaces of revolution). The physical layers, which are deployed
along these helicoids, have to be parallel in order to minimize compression
energy. We are thus led to study sets of parallel general helicoids, and
their helices.

The property of columns equidistance, which is preserved in each individual
helicoid, is still true for columns belonging to different parallel
helicoids. In effect, consider two neighboring parallel surfaces H and H' at
a distance $\delta \lambda $ measured along their common normals. Let%
\begin{equation}
G=\sigma _{1}\sigma _{2},\text{ \ \ \ \ \ \ \ \ \ \ \ \ \ \ \ \ \ \ \ \ \ \
\ \ \ \ \ \ \ \ \ \ \ \ }H=\frac{1}{2}(\sigma _{{\rm 1}}+\sigma _{{\rm 2}})
\label{eq:CurvH}
\end{equation}%
be the Gaussian and mean curvatures at some point M on H. $G$ and $H$ are
constant along the helix h passing through M on H, and depend upon the
unique parameter that defines the helix. Two parallel surfaces have the same
centers of curvature C$_{1}$ and C$_{2}$ at corresponding points, say M (on
H) and M' (on H'). Therefore%
\begin{equation}
\sigma _{{\rm 1}}^{\prime }=\frac{\sigma _{{\rm 1}}}{1+\delta \lambda \sigma
_{1}}\text{ , \ \ \ \ \ \ \ \ \ \ \ \ \ \ \ \ \ }\sigma _{{\rm 2}}^{\prime }=%
\frac{\sigma _{2}}{1+\delta \lambda \sigma _{2}}  \label{eq:sigma2}
\end{equation}

Thus the Gaussian and mean curvatures $G$' and $H$' are constant along the
curve h' drawn on the surface H' by the normals leaning on the first helix
h. Therefore H' is invariant under a one-parameter family of bendings; it is
indeed a helicoid, as already stated. On the other hand, h', which by
construction is at a constant distance $\delta \lambda $ of h, is an helix
on H', and thus belongs to a family of equidistant curves on H', all
obtained by tracing the curves h'($\delta \lambda $) equidistant to the
helices h.

This property explains why the isometric principle is satisfied not only on
each helicoid separately, but also throughout the whole HR. Notice that the
columns are not parallel, but anyway equidistant, which allows the
modulations to fit tightly from one layer to the next. It is fully evident
that a tight fitting is not feasible with planar parallel layers.

\subsection{Analytical Description of a Simple Stacking of Helicoids,
Ingredients}

A parallel stacking of helicoids on a ruled helicoid has been considered in %
\cite{MKleman3}. We revisit the corresponding analytical description, but
the detailed calculations, already in \cite{MKleman3}, will not be given.
The next section will be devoted to the calculation of the energy of such an
assembly of layers, and its comparison with a cylindrical stacking, which
yields a somewhat comparable but in fact fundamentally different assembly of
helical columns.

This subsection gives the analytical expressions necessary to calculate the
energy of a HR (next section); it is mostly technical, and can be skipped in
a first reading. Note however the equations \ref{eq:H0} and \ref{eq:Hlam},
for the notations which are used; they give the expressions of the central
ruled helicoid H(0) and of any parallel helicoid H($\lambda $) at a distance 
$\lambda $. Equations \ref{eq:npolar} and \ref{eq:ncartes} give the
coordinates of the tangent unit vector $\overrightarrow{n}$ along the
helices (the columns); the important point is that $\overrightarrow{n}$ is
rotationally invariant about the $z$-axis, which lets suppose that the
description of a HR in terms of nested cylinders is as valid a description
as in terms of nested helicoids. In fact, it is not so; we will discuss
later on why.

Let $\lambda $ be the signed distance of the helicoid H($\lambda $) to the
central right helicoid H(0). A point M on H($\lambda $) is identified by the 
$r$ and $\theta $ values of the foot on H(0) of the normal to H($\lambda $)
in M. Because the helicoids are parallel, they all have the same set of
normals. The coordinates of a point on H(0) are%
\begin{equation}
H(0)\equiv \{x=r\cos \theta ,y=r\sin \theta ,z=\varpi \theta \};\text{ }
\label{eq:H0}
\end{equation}

those of a point on H($\lambda $) are 
\begin{eqnarray}
H(\lambda ) &\equiv &\{x=r\cos \theta +\frac{\lambda \varpi }{N}\sin \theta
,y=r\sin \theta -\frac{\lambda \varpi }{N}\cos \theta ,z=\varpi \theta +%
\frac{\lambda r}{N}\},  \label{eq:Hlam} \\
N^{2} &=&r^{2}+\varpi ^{2}\text{ .}  \label{eq:N2}
\end{eqnarray}

We introduce the notation $t=N^{2}/\lambda \varpi $. The Gaussian curvature
and the mean curvature at M can be written%
\begin{eqnarray}
G &=&\sigma _{1}\sigma _{2}=-\frac{t^{2}\varpi ^{2}}{(t^{2}-1)N^{4}}=-\frac{1%
}{\lambda ^{2}(t^{2}-1)},  \label{eq:Gtlam} \\
H &=&\frac{1}{2}(\sigma _{{\rm 1}}+\sigma _{{\rm 2}})=\frac{\lambda
t^{2}\varpi ^{2}}{(t^{2}-1)N^{4}}=\frac{1}{\lambda (t^{2}-1)}
\label{eq:Htlam}
\end{eqnarray}

One observes that:

(1)- the mean curvature vanishes for $\lambda =0$, {\it i.e}. on the central
helicoid H(0), Fig. 5b. H(0) is indeed a (ruled) {\it minimal} surface. Note
that the sign of the mean curvature depends on the orientation of the normal
to H($\lambda $); here we have assumed that the three vectors $%
\overrightarrow{\frac{\partial M}{\partial r}}$, $\overrightarrow{\frac{%
\partial M}{\partial \theta }}$, and the normal $\overrightarrow{\nu }$ to H(%
$\lambda $) in M form a right-handed frame of reference.

(2)-the Gaussian curvature is negative for $(t^{2}-1)>0$, positive
otherwise, and is infinite for $t=\pm 1$. The region of negative Gaussian
curvature contains the central helicoid -- since it has negative Gaussian
curvature ($t=\pm \infty $) -- and extends symmetrically on both sides of
H(0) until it meets the two focal surfaces ($t=\pm 1$) of the family of
helicoids, i.e., the envelop of their normals. These focal surfaces were
studied in some details in \cite{MKleman3}; see also \cite%
{MKlemanLavrentovich}; we recall that each of them possesses a {\it helical}
singular line, around which the layers show a $k=+1/2$ configuration.

(3)-$G$, $H$, and $\sigma _{1},\sigma _{2}$ depend only on $r$ (not on $%
\theta $){\it \ i.e.} on the parameter that is a constant on a helix.
Therefore the curvatures properties, in particular the Gaussian curvature,
are constant along a given helix, which agrees with our analysis above.

The coordinates of the tangent unit vector along the helices can be written:%
\begin{equation}
\overrightarrow{n}\equiv \{\cos \theta \cos \varphi -\frac{r}{N}\sin \theta
\sin \varphi ,\sin \theta \cos \varphi +\frac{r}{N}\cos \theta \sin \varphi ,%
\frac{\varpi }{N}\sin \varphi \};\text{ }  \label{eq:npolar}
\end{equation}

$\varphi $ given by $t=\tan \varphi =N^{2}/\lambda \varpi .$

\qquad Using Eq.\ref{eq:Hlam}, $\overrightarrow{n}$ can also be written%
\begin{equation}
\overrightarrow{n}\equiv \frac{\sin \varphi }{N}\{-y,x,\varpi \}.\text{ }
\label{eq:ncartes}
\end{equation}

Hence $\overrightarrow{n}$ is a splayless ($div\overrightarrow{n}=0$)
double-twisted director field, rotationally symmetric about the $z$-axis.
The expressions below will be used in evaluating the energy of the system:%
\begin{eqnarray}
\overrightarrow{n}\cdot curl\overrightarrow{n} &\equiv &-2\varphi ^{\prime },%
\text{ \ \ \ \ \ \ \ \ \ \ \ \ \ \ \ \ \ \ \ \ \ \ \ \ \ \ \ \ \ \ \ \ \ \ \
\ \ \ \ \ \ }\left| \overrightarrow{n}\times curl\overrightarrow{n}\right|
^{2}=-\frac{1}{\varpi }\varphi _{\lambda }^{\prime }-\varphi _{\lambda
}^{\prime 2},\text{ }  \label{eq:curl2} \\
\overrightarrow{n}\times curl\overrightarrow{n} &=&\frac{1}{N}\sin \varphi
\left\{ n_{y},-n_{x},0\right\} ,\text{ \ \ \ \ \ \ \ \ \ \ \ \ \ \ \ \ \ \ \ 
}\left| curl\overrightarrow{n}\right| ^{2}=-\frac{1}{\varpi }\varphi
_{\lambda }^{\prime }+3\varphi _{\lambda }^{\prime 2},\text{ }  \nonumber
\end{eqnarray}

where $\varphi _{\lambda }^{\prime }=-\frac{1}{2\lambda }\sin \varphi $.\
The first equation on the second line requires $-1<\varpi \varphi _{\lambda
}^{\prime }<0$. In the applications, we shall assume, without loss of
generality, $\varpi >0$, $\varphi _{\lambda }^{\prime }<0$. Thus $\sin
2\varphi >0$, and $0<\varphi <\pi /2$, but because $\left| \tan \varphi
\right| >1$, one gets: $\pi /4<\varphi <\pi /2$. Furthermore, because of the
symmetry $\lambda \leftrightarrow -\lambda $, the points M$_{+}$ on H(+$%
\lambda $) and M$_{-}$ on H(--$\lambda $) are coupled by pairs that are
symmetric with respect to the intersection of the segment M$_{+}$M$_{-}$with
H(0) in its middle; we restrict in the sequel to the region $\lambda \geq 0$.

There are two types of helicoids H($\lambda $), those that are everywhere of
negative Gaussian curvature ($\lambda <\varpi $), those whose Gaussian
curvature changes sign ($\lambda >\varpi $). The sign of the Gaussian
curvature is negative if $t^{2}>1,$\ which can also be written $%
r^{2}>-\varpi ^{2}+\lambda \varpi $.

(1) - {\it Helicoids} $\lambda <\varpi $

\begin{equation}
0<r<\infty ;\text{ \ }\arctan \frac{\varpi }{\lambda }\text{ 
\mbox{$<$}%
}\varphi <\frac{\pi }{2}.  \label{eq:r}
\end{equation}

These helicoids do not touch the focal surfaces. They are folded around the
helical disclination $k=1/2$ that corresponds to the helix $\varphi =\pi /4$%
, $r=0$, $\left\{ x=\varpi \sin \theta ,y=-\varpi \cos \theta ,z=-\varpi
\theta \right\} $ on the helicoid H$(\lambda =\varpi )$. The Gaussian
curvature is negative over the entire helicoid, when $\lambda <\varpi $.

(2) - {\it Helicoids} $\lambda \geq \varpi $

One distinguishes two regions

(a) - negative Gaussian curvature:

\begin{equation}
\varpi \left( \lambda -\varpi \right) <r^{2}<\infty ;\text{ \ }\frac{\pi }{4}%
\text{ 
\mbox{$<$}%
}\varphi <\frac{\pi }{2}.  \label{eq:r2a}
\end{equation}

(b)- positive Gaussian curvature%
\begin{equation}
0<r^{2}<\varpi \left( \lambda -\varpi \right) ;\text{ \ }\arctan \frac{%
\varpi }{\lambda }\text{ 
\mbox{$<$}%
}\varphi <\frac{\pi }{4}.  \label{eq:r2b}
\end{equation}

\subsection{Why Negative Gaussian Curvature}

The helicoids are never physically present in their entirety in a HR; and
the physical parts of the helicoids are those which have negative Gaussian
curvature. The regions with positive Gaussian curvature are disfavored, and
the span of the helicoids is finite, for the following reasons:

(a)- The helicoids $\lambda <\varpi \ $(which have $G<0$) are free of any
singularity. The helical disclination $k=1/2$ , which is the first
singularity met when proceeding away from the central axis, belongs to the
helicoids $\lambda =\varpi $. But the helicoids $\lambda <\varpi $ do not
extend very far from the central axis because, as in double-twisted
cylinders in blue phases, the bending energy of the columns (energy ruled by
the Frank constant $K_{3}$) becomes prohibitive for large radii. Which
distance $l$ from the axis is difficult to guess at this stage; in analogy
with BP's, one could take $l$ $\cong $ $\varpi ,$ which yields $r_{\max
}\left( \lambda \right) ^{2}\cong \varpi \sqrt{\varpi ^{2}-\lambda ^{2}}$ ($%
r_{\max }(\lambda )$ is not the distance of a helix to the axis, but the
span of the helicoid; see equation \ref{eq:Hlam} and appendix). On the other
hand the observations reported in \cite{YuNastishin} indicate that the
domain of existence of the HR is not limited by a cylinder; taking then $%
r_{\max }\cong \varpi $ (independent of $\lambda $), one gets $l(\lambda )=%
\sqrt{\varpi ^{2}+\frac{\lambda ^{2}}{2}}$. Equation \ref{eq:r} is then
replaced by%
\begin{equation}
0<r<r_{\max }=\varpi ;\text{ \ }\arctan \frac{\varpi }{\lambda }\text{ 
\mbox{$<$}%
}\varphi <\arctan \frac{2\varpi }{\lambda }.  \label{eq:rmax}
\end{equation}

(b)- A solution with positive Gaussian curvature has an obvious
disadvantage: because the columns have a finite width, a whole number of
columns does not fill the interior of a layer without void or overlapping;
and in case it succeeds to fill the interior of some layer without void or
overlapping, it does not in the next one. Therefore the columnar packing
necessarily exhibits edge dislocations, whose energy will be estimated later
on to be much larger than the energy of the helical geometry. Contrariwise,
the columns have space to move locally along a lateral direction if the
layers have Gaussian negative curvature, and in fact to take in the layers a
distance of equilibrium that can be different from the distance from a layer
to the next, without necessitating edge dislocations. Thus, parallel
helicoids are preferred to nested cylinders, although the distribution of
the columns is rotationally invariant (in a continuous analytical
description).

\section{The Free Energy}

As already stated, the ribbons, according to \cite{YuNastishin}, are
characteristic of {\it metastable} phases B7$^{\text{*}}$; which in addition
to Sm and D orders, display helicity (H order), either left or right. Thus
the finite pitch $p$ $(=2\pi /q)$ of the columns is a material constant
(which might depend on the temperature). Assuming this to be true, we show
now, in the framework of a very simple model, that $\varpi q\cong -1$, {\it %
i.e.} $b\cong -p$.

The free energy of a HR contains a part relative to the columns, with
density $e_{ch}$, total energy $E_{ch}$, which can be written as the free
energy of a cholesteric phase oriented along $\overrightarrow{n}$, and a
part relative to the layers, layer energy density $e_{lay}$, total energy $%
E_{lay}$, which can be written as the free energy of a smectic phase. We
shall assume that the phenomenological free energy density is a sum of these
two contributions, without any term of interaction. All the interaction is
indeed contained in the isometric geometry requirements: the columns are
equidistant and the layers are parallel. For the same reason, we do not
introduce any energy of elastic (positional) deformation.

The helicity of a HR is described by two pseudoscalars

(a) - the Burgers vector $b$ \ of the dislocation, $b=nd_{0},n\in Z$, $d_{0}$
is the smectic repeat distance. We take $b$ positive if the dislocation is
right screw, negative in opposite case; $\varpi =b/2\pi $.

(b) - the pitch of the the twisted columns. This pitch is perpendicular to
the layers. We take the pitch positive if the helicity of the columns is
right handed, negative in opposite case; $q=2\pi /p$.

The 'cholesteric part' of the energy reads

\begin{equation}
e_{ch}=\frac{1}{2}K_{2}\left( \overrightarrow{n}\cdot curl\overrightarrow{n}%
+q\right) ^{2}+\frac{1}{2}K_{3}\left( \overrightarrow{n}\times curl%
\overrightarrow{n}\right) ^{2}=\left[ 2K_{2}-\frac{1}{2}K_{3}\right] \varphi
_{\lambda }^{\prime 2}-2\left[ qK_{2}+\frac{1}{4\varpi }K_{3}\right] \varphi
_{\lambda }^{\prime }+\frac{1}{2}K_{2}q^{2}  \label{eq:ech}
\end{equation}

Observe tat if $\varphi _{\lambda }^{\prime }>0$, one expects $q>0$, because
the twist $q$ contributes to a reduction of the energy density. Therefore,
according to our discussion of the signs above, the inverse pitch $q$, and
the Burgers vector $b$ (which is the pitch of the helicoids) are of opposite
signs. At this level of calculation we do not specify the origin of $q$ when
assigning a cholesteric energy density to the director field of the B7$^{%
\text{*}}$ phase. In other words, here $q$ is not necessarily a material
constant (although such a possibility is not excluded) but rather an
empirical model parameter that reflects the (metastable) helicity of the
columns in the B7$^{\text{*}}$ HRs. The result given below shows that q is
intimately related to the Burgers vector $b$.

The total `cholesteric' energy per unit length along the axis is

\begin{equation}
E_{ch}=\frac{1}{b}\int \int \int e_{ch}d\lambda dS\text{, \ \ \ \ }dS=\frac{%
t^{2}-1}{t^{2}}Ndrd\theta ;  \label{eq:Ech}
\end{equation}

$dS$ is the element of area of H($\lambda $). When integrating along one
pitch of the helicoids, $\theta $ runs over the interval $[0,2\pi \lbrack $;
it contributes by a factor $2\pi $ to the integration, since no term depends
on in $\theta $\ the integrand. The rest of the region of integration is a
variable of the problem, which is adjusted by the competition between layer
curvature effects (discussed below) and cholesteric contributions. In this
section, we just assume that it scales like $\varpi ^{2}$. This region is
necessarily bound by the two sheets of the focal surface F (the loci of the
centers of curvature of the helicoids) on which $\tan 2\varphi =1$. Starting
from the central helicoid H($\lambda =0$), the first helicoids for which
this equality is obtained are H($\lambda =\varpi $) and H($\lambda =-\varpi $%
), at $r=0$ (i.e. on the straight generators of H($\lambda =0$), at a
distance $\pm \varpi $ of the central axis). For the time being, and in
order to avoid a difficult analytical calculation, we write $E_{ch}\approx
\left\langle e_{ch}\right\rangle \varpi ^{2}$, with $\left\langle \varphi
_{\lambda }^{\prime }\right\rangle =-\frac{1}{2}\varpi $, $\left\langle
\varphi _{\lambda }^{\prime 2}\right\rangle =\frac{1}{4}\varpi ^{2}$,{\it \
i.e}.

\begin{equation}
E_{ch}=\left\langle e_{ch}\right\rangle \varpi ^{2}=\frac{1}{2}K_{2}+\frac{1%
}{8}K_{3}+K_{2}q\varpi \left[ 1+\frac{1}{2}q\varpi \right]  \label{eq:Ech2}
\end{equation}

If this expression were the only energy term, one would get, by minimizing
with respect to $\eta =\varpi q$:

\begin{equation}
\varpi q=-1  \label{eq:omegaq}
\end{equation}

The sign of the $\varpi $ and of the $q$ are opposite. It is easy to check
that $\left\langle e_{ch}\right\rangle $ is a minimum for this value of $%
\varpi $ ({\it i.e.} $\frac{\partial ^{2}\left\langle e_{ch}\right\rangle }{%
\partial \varpi ^{2}}>0$). The contribution of the layer curvature energy
does not change much these results, as we show now.

The layer curvature of the energy density reads:

\begin{equation}
e_{lay}=2\kappa \left( H\right) ^{2}+\kappa _{b}G  \label{eq:elay}
\end{equation}

$\kappa $ and $\kappa _{b}$ are referring to one layer. The total energy can
be written:

\begin{equation}
E_{lay}=\int \int \int e_{lay}\frac{dSd\lambda }{d_{0}}\text{,}
\label{eq:Elay}
\end{equation}

$d_{0}$ being the layer thickness. We discuss the $\kappa $ and $\kappa _{b}$
term separately.

{\it Mean curvature}. The set of stacked layers is built upon a central
layer that is a minimal surface. This situation is most probably generic,
because it yields such a small mean curvature energy. We shall therefore
neglect the `regular' part of this energy, keeping only the `singular' part,
that originates in the presence of the $k=1/2$ singularity at $r=0$, $%
\lambda =\varpi $. The disclination length is $b\sqrt{2}$ per turn. We thus
expect an energy of the order of: $\frac{\pi \sqrt{2}k^{2}\kappa }{d_{0}}\ln 
\frac{\varpi }{d_{0}}$ per unit length of helical axis.

{\it Gaussian curvature}. Assume that in each layer the boundaries of the
domain of existence of the phase are along two columns $r_{1}$, $r_{2}$.
According to the Gauss-Bonnet theorem we have $\int \int GdS=\pi -\int
\sigma _{g}dS$, with $\sigma _{g}$ the geodesic curvature. The geodesic
curvature vanishes for a geodesic, and takes the value $\sigma _{g}=\pm 
\frac{r}{\lambda \varpi }\cos \varphi =\pm \frac{r}{N^{2}}\sin \varphi $ on
a helix (a column), see appendix. The length of a column parameterized by $r$
is $1/\sin \alpha $ per unit length of helical axis, $\alpha \left( r\right)
\ $being the angle of the tangent $\overrightarrow{n}$ to the column with
the plane orthogonal to the HR axis; $\sin \alpha =n_{z}=\frac{\varpi }{N}%
\sin \varphi $, which is positive according to our conventions. We thus
expect an energy of the order of $\kappa _{b}\left[ \frac{\pi }{b}-\frac{%
r_{1}}{\varpi N_{1}}+\frac{r_{2}}{\varpi N_{2}}\right] $per unit length of
helical axis for the layer H($\lambda $). Because $N^{2}=r^{2}+\varpi ^{2}$
does not depend on $\lambda $, the contribution is the same for each layer,
and the expression for one layer has to be multiplied by the number of
layers $\varpi /d_{0}$. Take $r_{1}=0$, $r_{2}=\varpi $, one gets $\kappa
_{b}\left[ \frac{\pi }{b}-\frac{r_{1}}{\varpi N_{1}}+\frac{r_{2}}{\varpi
N_{2}}\right] =\kappa _{b}\pi \frac{1+\sqrt{2}}{b}$. Observe however that
the contribution of the disclination is certainly much larger than the
contribution of the Gaussian curvature, the ratio being of the order of $%
\kappa \varpi /\kappa _{b}d_{0}$, where $\left| \frac{\kappa }{\kappa _{b}}%
\right| \cong 1$, and $\frac{\varpi }{d_{0}}>>1$. We therefore neglect the $%
\kappa _{b}\ $contribution and eventually get:

\begin{equation}
E_{lay}\cong \frac{\pi \sqrt{2}k^{2}\kappa }{d_{0}}\ln \frac{\varpi }{d_{0}}=%
\frac{\pi \sqrt{2}\kappa }{4d_{0}}\ln \frac{\varpi }{d_{0}}  \label{eq:Elay2}
\end{equation}

This exspression is of the same order of magnitude as, althoug presumably
smaller than (see below, after Eq.\ref{eq:inequa}) $E_{ch}$, the ratio being
of the order of $E_{ch}/E_{lay}\cong K_{2}d_{0}/\kappa $.

We now minimize the total energy $E=E_{lay}+E_{ch}$ with respect to the
variable $\eta =q\varpi $. One gets:

\begin{equation}
\frac{\partial E}{\partial \eta }=\frac{\pi \sqrt{2}\kappa }{4d_{0}}\frac{1}{%
\eta }+K_{2}\left( 1+\eta \right) \text{; \ \ \ }\frac{\partial ^{2}E}{%
\partial \eta ^{2}}=-\frac{\pi \sqrt{2}\kappa }{4d_{0}}\frac{1}{\eta ^{2}}%
+K_{2}\text{\ \ }  \label{eq:dE}
\end{equation}

the analysis of the stability of the solution (which obeys $\frac{\partial E%
}{\partial \eta }=0$), starts with the condition $\frac{\partial ^{2}E}{%
\partial \eta ^{2}}>0$, in which inequality one inserts the value of $\eta $
for which $\frac{\partial E}{\partial \eta }=0$. After some algebra, one
finds one stable solution only

\begin{equation}
\varpi q=-\frac{1}{2}-\frac{1}{2}\sqrt{1-\pi \sqrt{2}\frac{\kappa }{%
K_{2}d_{0}}}  \label{eq:omegaq2}
\end{equation}%
that gives back equation for $\kappa =0$. Notice that the existence of such
a solution requires that

\begin{equation}
1-\pi \sqrt{2}\frac{\kappa }{K_{2}d_{0}}>0  \label{eq:inequa}
\end{equation}%
{\it i.e}. that $\kappa /d_{0}$ be smaller than $K_{2}$. This seems to be
generically the case. One can indeed argue as follows. Due to the anisotropy
of the splay modulus, the layers most certainly fold more easily about the
columns than perpendicularly to the columns. Call $\sigma _{p}$ the
curvature of a planar section of H($\lambda $) perpendicular to the columns (%
$\sigma _{p}$ measures the folding of the layer about the columns), $\sigma
_{c}$ the curvature of a section parallel to the columns ($\sigma
_{p}+\sigma _{c}=2H$), one finds

\[
\frac{\sigma _{p}}{\sigma _{c}}=-1+2\frac{1+t^{2}}{1-t^{2}}\text{, \ \ \ \ }%
t^{2}>1\text{,} 
\]%
so that $\left| \frac{\sigma _{p}}{\sigma _{c}}\right| >3$. Therefore the
layers are essentially folded about the columns, and the splay modulus at
work is small. Furthermore $K_{2}$, which is the twist constant of long (in
principle infinite) columns, is expected to be large. Therefore the
inequality \ref{eq:inequa} is certainly satisfied, and $\varpi q$ of Eq.\ref%
{eq:omegaq2} is a stable solution.

It remains to show that the above solution is much less costly than a
solution with circular cylindrical layers, which would necessarily contain
edge dislocations of the columns (with Burgers vectors $d_{c}$, which is the
columnar repeat distance. The density of such dislocations would be $\rho
_{e}\approx 1/Rd_{c}$, at a distance $R$ from the cylinders axis. Therefore
the number of dislocations in a cylinder of radius $\varpi $ (which scales
like their length per unit length of cylinder) would be $\rho _{e}\approx
\int RdRd\theta /rd_{c}=2\pi \varpi /d_{c}$. The associated energy is of the
order of $Bd_{c}^{2}$ per unit length of dislocation, {\it i.e}., $%
E_{disl}\approx 2\pi Bd_{c}\varpi $ per unit length of cylinder. Comparing
with the energy of the HR calculated above, one gets $E_{HR}/E_{disl}\approx 
\frac{K_{2}}{B}\frac{1}{d_{c}\varpi }$. In this expression, $K_{2}/B$ is a
squared microscopic length, of the same order as $d_{c}^{2}$. Therefore $%
E_{HR}/E_{disl}$ is the ratio of a microscopic length over a macroscopic
length, and is very small.

\section{Discussion}

Isometric textures are the rule in liquid crystalline phases as soon as
there is one or two dimensions of periodicity \cite{MKlemanLavrentovich}:
focal conic domains when the order parameter is layer-like (\{Sm\}, one
dimension of periodicity), developable domains when it is columnar (\{D\},
two dimensions of periodicity). In both cases the associated distorted
textures are metastable and carry singularities. The situation is more
involved and really fascinating when is added to the periodicities a trend
to helicity, \{H\}. Twist grain boundary phases (TGB) \cite{Renn} and Blue
Smectic phases (BPSm) \cite{Pansu} offer remarkable examples of the \{Sm+H\}
mixed case; in both cases the helicity axis is in the layer and affects the
molecules. Then, one gets frustrated phases, regular, periodic, with
thermodynamically stable singularities (dislocations, mostly). \{H+D\}
frustrated phases have been suggested in \cite{Kamien}; they could display
domains of parallel columns separated by twist walls, possibly observed in %
\cite{Leforestier}.

The \{Sm+D\} order parameter is such that the layers are modulated in
thickness and fit closely along the modulations. We have demonstrated that
helical ribbons are manifestations of the isometric principle for this order
parameter. Those HRs have a finite width and do not transform in another
texture when they come into close contact. For instance, they do not extend
laterally into quasi-planar layers, after the manner of the giant screw
dislocations in SmAs, alluded to above. This extension would require a
slight breaking of isometry, and the columns would no longer be twisted,
just bent. It is probably this latter character that forbids such an
extension. In other words the helicity seems to belong to the order
parameter, which is therefore \{Sm+D+H\}. But there is no way to construct
in Euclidean space an infinite crystal with helicity showing the same close
packing as in the \{Sm+D\} case; the HRs can then be said to be frustrated
textures. The \{Sm+D+H\} phase is most probably crystal-definable in a
curved space, but we do not have investigated this question. Note however
that the isometric principle could be directly applied to this curved
crystal, and would then yield the HRs.

It is worth mentioning another character which differentiates SmAs giant
screw dislocations and HRs. While the HR geometry defined in section II is
experimentally strictly obeyed (we have already mentioned the rigidity of
these textures), the inner geometry (inside the helical disclinations) of a
Sm screw dislocation is very flexible and susceptible of many deformations.
In fact, the giant screw dislocation is only an approached isometric texture
for a Sm phase (the true isometric textures are focal conic domains), and
can therefore suffer large fluctuations, in particular due to the presence
of edge and screw dislocations of small Burgers vectors (\cite{KlemanPLW},
chapter 5).

We have mentioned the analogy between a HR and a double-twist BP cylinder.
In both cases the columns (in the BP case what we call columns are the
envelops of the director field) are double-twisted and rotationally
invariant about the axis direction. Because BP is a true 3D liquid, there is
no obstacle to merging when two double-twist cylinders come into contact;
the director field deforms without much energy expense between cylinders,
and the cylinder geometry fills space \cite{Meiboom}. Now, as already
stated, a layer helical texture is preferable to a layer cylindrical texture
for a HR. There are no layers in a BP, and it is the custom to desribe this
double-twist texture as if the columns are partitioned into cylinders. It is
true that isometry requirements are weak in BP, as long as the molecules
forming the columns are short compared to the pitch. However, it is believed
that BPs are favored with long molecules \cite{Grebel,Kleman5}, and
double-twist cylinders are also present in certain DNA solutions \cite%
{Livolant2}. One might therefore wonder whether a partition into parallel
helicoids (compared to nested cylinders, this requires a small displacement
of the columns) might not be closer to reality in some BPs.

A final, brief, remark relating to the HR width limitation. Let us notice
that the region of negative Gaussian curvature is bound by the focal surface
of the stacking, whose two sheets obey the equation $t=\pm 1$, see section
II, C. This is a first reason why the width stops short of the focal
surface. The helical disclinations belong to the focal surfaces, which
extend outwards. A second reason relates to the twist of the columns:
notice, equation \ref{eq:curl2}, that $\overrightarrow{n}\cdot curl%
\overrightarrow{n}\sim 1/\lambda $ for $\lambda $ large, i.e. the twist
vanishes for the helicoids that are far enough from the axis; consequently
the twist energy becomes large for $\lambda $ large, if, as assumed, $p$ is
a material constant (metastability of the B7$^{\text{*}}$ phase); at the
same time the bend energy decreases. Observe that these behaviors of the
twist and the bend energy for $\lambda $ large are both opposite to their
behaviors for the cylinders of double twist. But in both cases one expects
that the competition between these two energy terms fixes the width of the
texture at some finite value which scales with the pitch.

{\bf Acknowledgement. }M.K. thanks Dr. Fran\c{c}oise Livolant for
discussions. Yu.N. acknowledges discussions with Dr. Ph. Barois and the
support and hospitality of CRPP, Universit\'{e} de Bordeaux-I.

\section{Appendix}

a) width of a HR.

Let $\tan \psi =\lambda \varpi /Nr$. Eq.\ref{eq:Hlam} can be written

\begin{equation}
H(\lambda )\equiv \{x=l\cos \left( \theta -\psi \right) ,\text{ \ }y=l\sin
\left( \theta -\psi \right) ,\text{ }z=\varpi \left( \theta -\psi \right)
+\eta \left( l\right) \},  \label{Hlam1}
\end{equation}

where $l^{2}=r^{2}+\left( \frac{\lambda \varpi }{N}\right) ^{2}$is the
distance to the $z$-axis; $\eta \left( l\right) =\varpi \psi +\frac{r^{2}}{%
\varpi }\tan \psi $.

Assume now that $r_{\max }=\varpi $, one gets:

\[
\tan ^{2}\psi =\frac{\lambda ^{2}}{2\varpi ^{2}};\ \ \ l_{\max }=\sqrt{%
\varpi ^{2}+\frac{\lambda ^{2}}{2}}=\frac{\varpi }{\cos \psi };\ \ \eta
\left( l_{\max }\right) =\varpi \psi +\frac{\lambda }{\sqrt{2}}. 
\]

Therefore the equation of the boundary of the nucleus in a plane $z=$%
constant (here this constant is taken equal to zero) is, in polar
coordinates:

\[
\rho ^{2}=\varpi ^{2}\left( 1+\theta ^{2}\right) \text{.} 
\]

b) geodesic curvature.

Let $M$ be a 2D manifold, M be a point on $M$ moving along a curve C, $s$
the abscissa along C, $\nu $ the normal in M to $M$, we have, according to
well-known formulae of surface theory:

\[
\sigma _{\nu }=\overrightarrow{\nu }\cdot \overrightarrow{\frac{\partial
^{2}M}{\partial s^{2}}}\text{ \ \ \ \ \ \ \ \ \ \ \ \ \ }\sigma _{g}=\left( 
\overrightarrow{\nu }\times \overrightarrow{\frac{\partial M}{\partial s}}%
\right) \cdot \overrightarrow{\frac{\partial ^{2}M}{\partial s^{2}}}\text{
.\ \ \ \ \ } 
\]

The normal curvature $\sigma _{\nu }$ is the curvatureof the section normal
to $M$ at M; its sign depends on the sign of $\overrightarrow{\nu }$. The
geodesic curvature $\sigma _{g}$ measures the twist of the curve in the
plane tangent to $M$ in M. It vanishes for C a geodesic line. Its sign
depends on the orientation of the curve.

These quantities are easily calculated for a helix on the helicoid H($%
\lambda $). We have $\overrightarrow{\frac{\partial M}{\partial s}}=%
\overrightarrow{n}$. Thus, $\frac{\partial ^{2}M}{\partial s^{2}}$ follows
from the expression of $\overrightarrow{n}$ by differentiating at constant $%
r $. The calculation yields:

\[
\overrightarrow{\frac{\partial ^{2}M}{\partial s^{2}}}=\frac{\sin \varphi }{N%
}\text{\ }\left\{ -n_{y},n_{x},0\right\} \text{ \ }(=curl\overrightarrow{n}%
\times \overrightarrow{n})\text{\ .\ \ } 
\]

Hence:

\[
\sigma _{\nu }=-\frac{1}{\lambda }\cos ^{2}\varphi \text{\ \ \ \ \ \ \ \ \ \
\ \ \ }\sigma _{g}=\frac{r}{\lambda \varpi }\cos \varphi \text{.\ \ \ \ \ } 
\]

\newpage

\begin{center}
{\large Figure Captions}
\end{center}

Figure 1.Edge dislocations in a bent crystal: (a) at random, (b)
polygonized; (c) curved layers in a SmA, infinitesimally small edge
dislocations, isometric deformation.

Figure 2. The banana-like chemical configuration of D14F3.

Figure 3. A slim helical ribbon (S-HR) observed under the optical microscope.

Figure 4. Experimental data relating to the (slim) Helical Ribbons: (a)-
distribution of the widths; note that the experimental accuracy is $\sim
1\mu m$, so that each box is $2\mu m$ wide; (b)- ratio $b/l$ for the same
experimental data; this ratio does not vary much when the width deviates
from its more probable value.

Figure 5. Geometry of a helical ribbon: (a)- nested helicoids stacked
parallel to a central ruled helicoid; the domain width is limited $\left|
r\right| <b/2\pi $ (see Eq. \ref{eq:Hlam}); the `holes' in the figure result
from the small number of helicoids chosen parallel to the central HR; (b)-
the $\lambda =0$ (central helicoids) and $\lambda =0.5$ helicoids, with the
modulations (the columns) along helical lines; $\left| r\right| <b/2\pi $.
In each figure the two helices are the singularities of the stacking
(disclinations $k=1/2$). The Burgers vector is assumed positive; the
helicoids are right-handed.

\end{document}